\documentstyle[preprint,aps]{revtex}
\begin{document}
\draft
\preprint{ PUPT-1775 SPIN-1998/12 IASSNS-HEP-98-22 }
\title{Hawking Radiation As Tunneling}
\author{Maulik K. Parikh}
\address{Joseph Henry Laboratories, Princeton University, 
Princeton, New Jersey 08544, USA}
\address{Spinoza Institute, University of Utrecht,
3584 CE Utrecht, The Netherlands}
\author{Frank Wilczek}
\address{School of Natural Sciences, Institute for Advanced Study, 
Princeton, New Jersey 08540, USA}

\maketitle

\def \root {\sqrt{M^2 - Q^2}}
\def \Root {\sqrt{2Mr - Q^2}}

\begin{abstract}
We present a short and direct 
derivation of Hawking radiation as a tunneling process, based
on particles in a dynamical geometry. The imaginary part of the action 
for the classically forbidden process is related to the Boltzmann
factor for emission at the Hawking temperature. Because
the derivation respects conservation laws, the exact
spectrum is not precisely thermal.  We compare and contrast the
problem of spontaneous emission of charged particles from a charged
conductor.

\end{abstract}

\def \d {\delta}
\def \del {\nabla}
\def \pl {\partial}
\def \f {\frac}
\def \lf {\left (}
\def \rt {\right )}
\def \w {\omega}
\def \eps {\epsilon}
\def \Mw2 {\lf M - \f{\scriptstyle \w}{2} \rt}

\section{Introduction}
Several derivations of Hawking radiation exist in the literature 
\cite{swh2,gh}. 
None of them, however, correspond very directly to one of the heuristic
pictures most commonly proposed to visualize
the source of the radiation,  as tunneling.  
According to this picture, the
radiation arises by a process similar to electron-positron pair
creation in a constant electric field.  The idea is that the energy of
a particle changes sign as it crosses the horizon, so that a
pair created just inside or just outside the horizon can
materialize with zero total energy, after one member of the pair has
tunneled to the opposite side.  

Here we shall show that this schematic can 
be used to provide a short, direct semi-classical derivation of black hole
radiance. In what follows, energy conservation plays a fundamental
role: one must make a transition between states with the same total
energy, and the mass of the residual hole must go down as it radiates.
Indeed, it is precisely the possibility of lowering the black hole mass
which ultimately drives the dynamics.
This supports the idea that, in quantum gravity, 
black holes are properly regarded as highly excited states.

Broadly speaking, there are two standard approaches to Hawking radiation.
In the first, one considers a collapse geometry. The response of external
fields to this can be done explicitly or implicitly by abstracting
appropriate boundary conditions. In the second, one treats the black hole
immersed in a thermal bath. In this approach, one shows that (in general,
metastable) equilibrium is possible. By detailed balance, this implies
emission from the hole. 
In both the standard calculations, the background geometry is considered
fixed, and energy conservation is not enforced during the emission process.

Here we will consider a hole in empty Schwarzschild space, but with a 
dynamical geometry so as to enforce energy conservation.
(Despite appearances, 
the geometry is not truly static,  
since there is no global Killing vector.)  
Because we are treating this aspect more realistically, 
we must -- and do -- find corrections
to the standard results. These become quantitively significant when
the quantum of radiation carries a substantial fraction of the 
mass of the hole.

\section{Tunneling}

To describe across-horizon phenomena, it is necessary
to choose coordinates which, unlike Schwarzschild coordinates, are not
singular at the horizon.  A particularly suitable choice is obtained by 
introducing a time coordinate,
\begin{equation}
t = t_s + 2 \sqrt {2 M r} + 2M \ln \f {\sqrt{r} - \sqrt{2M}}
{\sqrt{r} + \sqrt{2M}} \; ,
\end{equation}
where $t_s$ is Schwarzschild time.  With this choice, the line element reads
\begin{equation}
ds^2 = - \lf 1 - \f{2M}{r} \rt dt^2 + 2 \, \sqrt{\f{2M}{r}} \,
 dt \, dr + dr^2 + r^2 d \Omega^2 \; .	\label{ds2}
\end{equation}
There is now no singularity at $r= 2M$, and the true character
of the spacetime, as being stationary but not static, is manifest. 
These coordinates were first introduced
by Painlev\'e \cite{pain} (who used them to criticize general
relativity, for allowing singularities to come and go!). Their utility
for studies of black hole quantum mechanics was emphasized more recently in
\cite{line}.

For our purposes, the crucial features of these coordinates are that they
are stationary and 
nonsingular through the horizon.  Thus it is possible to define an effective 
``vacuum'' state of a quantum field by requiring that it annihilate
modes which carry negative frequency with respect to $t$; such a
state will look essentially empty (in any case, nonsingular) 
to a freely-falling observer as he or she
passes through the horizon. This vacuum differs strictly from the
standard Unruh vacuum, defined by requiring positive frequency with
respect to the Kruskal coordinate $U = - \sqrt{r - 2M} \exp \lf -{t_s - r
\over 4 M} \rt$ \cite{unruh}. The difference, however, shows up only in
transients, and does not affect the late-time radiation.  

The radial null geodesics are given by
\begin{equation}
\dot{r} \equiv \f{dr}{dt} = \pm 1 - \sqrt{\f{2M}{r}} \; ,	\label{null}
\end{equation}
with the upper (lower) sign in Eq. \ref{null} 
corresponding to outgoing (ingoing) geodesics, under the implicit 
assumption that $t$ increases towards the future. 
These equations are modified 
when the particle's self-gravitation is taken
into account. Self-gravitating shells
in Hamiltonian gravity were studied by Kraus and Wilczek \cite{per}. They 
found that, when the black hole mass is held fixed and the total ADM mass
allowed to vary, a shell of energy $\w$ moves in the geodesics of a spacetime
with $M$ replaced by $M+\w$. If instead
we fix the total mass and allow the hole mass to fluctuate, 
then the shell of energy $\w$ travels on the 
geodesics given by the line element
\begin{equation}
ds^2 = - \lf 1 - {2 ( M - \w ) \over r} \rt dt^2 
+ 2 \, \sqrt{ {2 (M - \w ) \over r}} \, dt \, dr + dr^2 + r^2 d \Omega^2 \; ,
\end{equation}
so we should use Eq. \ref{null} with $M \to M - \w$.

Since the typical wavelength of the radiation
is of the order of the size of the black hole, one might doubt whether
a point particle description is
appropriate. However, when the outgoing wave is traced back towards the
horizon, its wavelength, as measured by local fiducial observers, is
ever-increasingly blue-shifted. Near the horizon, the radial wavenumber
approaches infinity and the point particle, or WKB,
approximation is justified.

The imaginary part of the action for an s-wave
outgoing positive energy particle  which crosses the horizon
outwards from $r_{\rm in}$ to $r_{\rm out}$ can be expressed as
\begin{equation}
{\rm Im}~ S = {\rm Im} \int_{r_{\rm in}}^{r_{\rm out}} p_r \, dr = 
{\rm Im} \int_{r_{\rm in}}^{r_{\rm out}} \! \!
\int_0^{p_r} d p'_r \, dr \; . \label{integrals}
\end{equation} 
Remarkably, this can be evaluated without entering into the details of
the solution, as follows.  We multiply and divide the integrand by the
two sides of Hamilton's equation
$\dot{r} = +\left. {dH \over d p_r} \right | _r \,$,  
change variable from momentum to
energy, and switch the order of integration to obtain
\begin{equation}
{\rm Im}~ S = {\rm Im} \int_M^{M - \w} \!\!
\int_{r_{\rm in}}^{r_{\rm out}} {dr \over \dot{r}} \, dH 
= {\rm Im} \int_0^{+\w} \!\!
\int_{r_{\rm in}}^{r_{\rm out}} {dr \over 1 - \sqrt {2 \lf M -
\w' \rt \over r}} \, \lf - d \w' \rt \; ,
\end{equation} 
where we have used the modified Eq. \ref{null},
and the minus sign appears because $H = M - \w'$. 
But now the integral can be done by deforming the contour, so as to
ensure that positive energy solutions decay in time (that is, into the
lower half $\w'$ plane). In this way we obtain
\begin{equation}
{\rm Im}~ S = + 4 \pi \w \lf M - {\w \over 2} \rt \; , \label{answer}
\end{equation} 
provided $r_{\rm in} > r_{\rm out}$. To understand this ordering --
which supplies the correct sign -- we observe that 
when the integrals in Eq. \ref{integrals} 
are not interchanged, and with the contour evaluated via the 
prescription $\w \to \w - i \eps$, we have
\begin{equation}
{\rm Im}~ S = + {\rm Im} \int_{r_{\rm in}}^{r_{\rm out}} \!\! 
\int_M^{M - \w} {d M' \over 1 - \sqrt{{2 M' \over r}}} \, dr
= {\rm Im} \int_{r_{\rm in}}^{r_{\rm out}} - \pi r \, dr \; .
\end{equation}
Hence $r_{\rm in} = 2 M$ and $r_{\rm out} = 2 \lf M - \w \rt$. 
(Incidentally, comparing the above equation with Eq. \ref{integrals}, 
we also find that ${\rm Im}~ p_r = - \pi r$.) Although this radially
inward motion appears at first sight to be classically allowed, it is
nevertheless a classically forbidden trajectory because 
the apparent horizon is itself contracting. Thus, the limits on
the integral indicate that, over the course of 
the classically forbidden trajectory, the outgoing particle starts from
$r = 2M - \eps$, just inside the {\em initial} position of the horizon, 
and traverses the contracting
horizon to materialize at $r = 2(M - \w ) + \eps$, just outside 
the {\em final} position of the horizon.

Alternatively, and along the same lines, Hawking
radiation can also be regarded as pair creation {\em outside\/} the horizon,
with the negative energy particle tunneling into the black hole. Since such
a particle propagates backwards in time, we have to reverse time in the
equations of motion. From the line element, Eq. \ref{ds2}, we see that
time-reversal corresponds to 
$\sqrt{\f{2M}{r}} \to - \sqrt{\f{2M}{r}}$. Also, since the anti-particle
sees a geometry of fixed black hole mass, the upshot of self-gravitation
is to replace $M$ by $M + \w$, rather than $M - \w$.
Thus an ingoing negative energy particle has
\begin{equation}
{\rm Im}~ S = {\rm Im} \int _{0} ^{- \w} \!\! \int _{r_{out}} ^{r_{in}} 
{dr \over -1 + \sqrt{{2 \lf M + \w' \rt \over r}} } \, d \w' 
= + 4 \pi \w \lf M - \f {\w}{2} \rt \; ,
\end{equation}
where to obtain the last equation we have used Feynman's ``hole theory'' 
deformation of the contour: $\w' \to \w' + i \eps$.

Both channels -- particle or anti-particle tunneling -- contribute to the
rate for the Hawking process so, in a more detailed calculation, one would
have to add their amplitudes before squaring in order to obtain the
semi-classical tunneling rate. Such considerations, however, only concern
the pre-factor.
In either treatment, the exponential part of the semi-classical emission
rate, in agreement with \cite{k5,serge}, is
\begin{equation}
\Gamma \sim e^{-2 \, {\rm Im}~ S} = e^{-8 \pi \w \Mw2 } = e^{+ \Delta 
S_{\rm B-H}} \; , \label{rate}
\end{equation}
where we have expressed the result more naturally in terms of the change 
in the hole's
Bekenstein-Hawking entropy, $S_{\rm B-H}$.
When the quadratic term is neglected, Eq. \ref{rate}
reduces to a Boltzmann factor for a particle with energy $\w$ at
the inverse Hawking temperature $8 \pi M$.  The $\w^2$ correction
arises from the physics of energy conservation, which (roughly
speaking) self-consistently raises the effective temperature of the
hole as it radiates.  
That the exact result must be correct can be seen on physical grounds by 
considering the 
limit in which the emitted particle carries away the entire mass and 
charge of the black hole (corresponding to the transmutation of the
black hole into an outgoing shell). There can be only one such outgoing 
state. On the other hand, there are $\exp \lf S_{\rm B-H} \rt$ states in total.
Statistical mechanics then asserts that the probability of finding a
shell containing all the mass of the black hole is proportional
to $\exp \lf - S_{\rm B-H} \rt$, as above.

Following standard arguments, Eq. \ref{rate} with the quadratic term
neglected implies
the Planck spectral flux appropriate to an inverse temperature of $8 \pi M$:
\begin{equation}
\rho \lf \omega \rt = \f{d \w}{2 \pi} \f{| \, T \lf \omega \rt | ^2} 
{e^{+ 8 \pi M \omega } - 1} \; ,
\end{equation}
where $| \, T \lf \omega \rt | ^2$ is the frequency-dependent (greybody) 
transmission co-efficient for the
outgoing particle to reach future infinity without back-scattering.
It arises from a more complete treatment of the modes, whose
semi-classical behavior near the turning point we have been discussing.

The preceding techniques can also be applied to emission from a charged black
hole. However, when the outgoing radiation carries away the 
black hole's charge, 
the calculations are complicated by the fact that the trajectories 
are now also subject to electromagnetic forces. 
Here we restrict ourselves to uncharged radiation coming from a 
Reissner-Nordstr\"om black hole. The derivation then proceeds in a 
similar fashion to that above.

The charged counterpart to the Painlev\'e line element is
\begin{equation}
ds^2 = - \lf 1 - {2M \over r} + {Q^2 \over r^2} \rt dt^2 + 2 \, 
\sqrt{{2M \over r} - {Q^2 \over r^2}} \,  dt \, dr + 
dr^2 + r^2 d \Omega^2 \; ,
\end{equation}
which is obtained from the standard Reissner-Nordstr\"om line element by 
the coordinate transformation,
\begin{eqnarray}
t & = & t_r + 2 \Root +M \ln \lf {r - \Root \over r + \Root} \rt \nonumber \\
& & \indent \indent + {Q^2 - M^2 \over \root} \, 
{\rm arctanh} \lf {\root \Root \over Mr} \rt \; ,
\end{eqnarray}
where $t_r$ is the Reissner time coordinate. The line element now 
manifestly displays the stationary, nonstatic, and nonsingular 
nature of the spacetime. 

The equation of motion
for an outgoing massless particle is
\begin{equation}
\dot{r} \equiv \f{dr}{dt} = + 1 - \sqrt{{2M \over r} - {Q^2 \over r^2}} \; ,
\end{equation}
with $M \to M - \w$ when self-gravitation is included \cite{percharged}. 
The imaginary part
of the action for a positive energy outgoing particle is
\begin{equation}
{\rm Im}~ S = {\rm Im} \int_0^{+\w} \!\!
\int_{r_{\rm in}}^{r_{\rm out}} {dr \over 1 - 
\sqrt {{2 \lf M - \w' \rt \over r} - {Q^2 \over r^2}} } \, \lf - d \w' \rt \; ,
\end{equation} 
which is again evaluated by deforming the contour in accordance with
Feynman's $w' \to w' - i \eps$ prescription. The residue at the pole
can be read off by substituting $u \equiv \sqrt{2 \lf M - \w' \rt r - Q^2}$. 
This yields an emission rate of
\begin{equation}
\Gamma \sim e^{-2 \, {\rm Im}~ S} 
= e^{- 4 \pi \lf 2 \w \Mw2 - (M- \w) \sqrt{(M-\w)^2 - Q^2} + M \root \rt}
= e^{+ \Delta S_{\rm B-H}} \; . \label{chargedgamma}
\end{equation}
To first order in $\w$, Eq. (\ref{chargedgamma}) is consistent with 
Hawking's result of
thermal emission at the Hawking temperature, $T_H$, for a charged black hole:
\begin{equation}
T_H = {1 \over 2 \pi} {\root \over \lf M + \root \rt ^2} \; .
\end{equation}
Again, energy conservation implies that the exact result has
corrections of higher order in $\w$; these can all be collected to 
express the emission rate as the exponent of the change in entropy. 
Moreover, since the emission rate has to be real, the presence of the 
first square root in Eq. (\ref{chargedgamma})
ensures that radiation past extremality is not possible.  Unlike in
the traditional formulas, the third law of black hole thermodynamics is 
here manifestly enforced.

Note that only local physics has gone into
our derivations. There was neither an appeal to Euclideanization nor any
need to invoke 
an explicit collapse phase. The time asymmetry leading to outgoing
radiation arose instead from use of the ``normal'' local contour deformation
prescription in terms of the nonstatic coordinate $t$.

\section{Relation to Electric Discharge}

The calculation presented above is formally self-contained, but some
additional discussion is in order, to elucidate its physical meaning
and to dispel a puzzle it poses.

When considered at the very broadest level, radiation of mass from a
black hole resembles tunneling of  electric charge off a charged
conducting sphere.    Upon a moment's reflection, however, the
difference in the physics of the two situations appears so striking as
to pose a puzzle.  For while the electric force between like charges
is repulsive, the gravitational force is always attractive.   Related
to this, the field energy of electric fields is positive, while
(heuristically) the field energy of gravitational fields is negative.
On this basis one might think that the electric process should proceed
spontaneously, and need not require tunneling, while the
gravitational process has no evident reason to proceed at all.

Consider a conducting sphere of radius $R$ carrying charge $Q$. The
electric field energy can be lowered by emitting a charge $q$
so we expect this process to occur spontaneously.   If we neglect
back-reaction of the charge $q$ on the conducting sphere, the force is
repulsive at all distances, and there is no barrier to emission.   In
a more accurate treatment, however, we must take into account the
induced non-uniformity of the charge on the sphere, which is easily
done using the method of images.   The effective potential is 
\begin{equation}
V(r) ~=~ q \lf {Q-q \over r} - {q R \over r^2 - R^2} \rt \; ,
\end{equation}
where we consider configurations of image charge which leave the
potential on the sphere constant and the field at infinity fixed.
In the formal limit $Q \gg q$ the first term dominates, and the
potential decreases monotonically with $r$, indicating no barrier.
However the second term increases monotonically with $r$, and always
dominates for $r \to R$, producing a barrier. 

In the gravitational problem, the situation is just the reverse.  With
back-reaction  neglected, there is nothing but barrier.  Yet our
calculation including back-reaction indicates the possibility of
redistributing mass-energy of the gravitating sphere (black hole) into
kinetic energy of emitted radiation.

Since the intrinsic energy of the gravitational field is negative,  it
is disadvantageous to reduce it, point by point.   However, since in general 
the spacetime containing a black hole is not globally static, there exist 
freely propagating negative energy modes inside the horizon which cause 
the black hole to shrink. As a consequence, the 
black hole's radius decreases and {\em the external volume
of space, over which one integrates the field, increases}.  This,
kinematically, is why the radiation process is allowed.  Were the hole
geometry to be regarded as fixed, there would be no possible source
for the kinetic energy of the radiation, and a genuine tunneling
interpretation of Hawking radiation would appear to be precluded.

\section{Conclusion}
We have derived Hawking radiation from the heuristically familiar perspective
of tunneling. Our derivation is in consonance with intuitive
notions of black hole radiance but, by
taking into account global conservation laws, we are led to a modification
of the emission spectrum away from thermality. The resulting
corrected formula has physically reasonable limiting cases and, by 
virtue of nonthermality, suggests
the possibility of information-carrying correlations in the radiation.

{\bf Acknowledgement}

F.W. is supported in part by DOE grant DE-FG02-90ER-40542.

\end{document}